\documentclass[11pt]{article}
\usepackage{a4wide}                      
\usepackage{epsf}                        
\usepackage{latexsym}                    
\usepackage{amsfonts}                    
\usepackage{amssymb}                     

\newcommand{\sect}[1]{\setcounter{equation}{0}\section{#1}}

\def\be{\begin{equation}}
\def\ee{\end{equation}}
\def\bea{\begin{eqnarray}}
\def\eea{\end{eqnarray}}

\def\nn{\nonumber \\ [.3cm]}

\def\hsp#1{\hspace{#1}}

\def\part{\partial}
\def\tfrac#1#2{{\textstyle{\frac{#1}{#2}}}}
\def\half{\tfrac{1}{2}}
\def\x{\times}

\def\incl{\mbox{i}}

\def\STr{\mbox{STr}}

\def\tB{\tilde B}

\def\hI{\hat I}

\def\hX{\hat X}
\def\hZ{\hat Z}

\def\dhX{\dot{\hat X}}

\def\hC{{\hat C}}

\def\ha{{\hat a}}
\def\hb{{\hat b}}
\def\hc{{\hat c}}
\def\hg{{\hat g}}
\def\hh{{\hat h}}
\def\hi{{\hat \imath}}
\def\hj{{\hat \jmath}}
\def\hk{{\hat k}}
\def\hl{{\hat l}}
\def\hm{{\hat m}}
\def\hn{{\hat n}}
\def\hp{{\hat p}}

\def\hmu{{\hat{\mu}}}
\def\hnu{{\hat{\nu}}}

\def\Ortin{Ort{\'\i}n}

\def\makeatletter{\catcode`\@=11}
\makeatletter
\def\mathbox#1{\hbox{$\m@th#1$}}%
\def\math@ccstyles#1#2#3#4#5#6#7{{\leavevmode
      \setbox0\mathbox{#6#7}%
      \setbox2\mathbox{#4#5}%
      \dimen@ #3%
      \baselineskip\z@\lineskiplimit#1\lineskip\z@
      \vbox{\ialign{##\crcr
             \hfil \kern #2\box2 \hfil\crcr
             \noalign{\kern\dimen@}%
             \hfil\box0\hfil\crcr}}}}
\def\mathaccstyles{\math@ccstyles\maxdimen}
\def\maththroughstyles{\math@ccstyles{-\maxdimen}}
\def\unity%
 {\maththroughstyles{.45\ht0}\z@\displaystyle {\mathchar"006C}\displaystyle 1}
\begin{document}

\rightline{DCPT-02/49}
\rightline{FFUOV-02/06}
\rightline{hep-th/0207199}
\rightline{15 January 2003}
\vspace{2cm}

\centerline{\LARGE \bf A Microscopical Description of Giant Gravitons}
\vspace{1.3truecm}

\centerline{
  {\large \bf Bert Janssen${}^{a,}$}\footnote{E-mail address:
    {\tt bert.janssen@durham.ac.uk} }
  {\bf and}
  {\large \bf Yolanda Lozano${}^{b,}$}\footnote{E-mail address:
    {\tt yolanda@string1.ciencias.uniovi.es}}
  }

\vspace{.8cm}
\begin{center}
  {\it ${}^a$ Department of Mathematical  Sciences, University of Durham, \\
    South Road, Durham DH1 3LE, United Kingdom}
\end{center}

\begin{center}
  {\it ${}^b$ Departamento de F{\'\i}sica,  Universidad de Oviedo, \\
    Avda.~Calvo Sotelo 18, 33007 Oviedo, Spain}
\end{center}
\vspace{2truecm}

\centerline{\bf ABSTRACT}
\vspace{.5truecm}

\noindent
We construct a non-Abelian world volume effective action for a system of multiple
M-theory gravitons. This action contains multipole moment couplings to the 
eleven-dimensional background potentials. We use these couplings to study, from the 
microscopical point of view, giant graviton configurations where the gravitons 
expand into an M2-brane, with the topology of a fuzzy 2-sphere, that lives in
the spherical part of the $AdS_7\times S^4$ background or in the $AdS$ part of
$AdS_4\times S^7$. When the number of gravitons is large we find perfect
agreement with the Abelian, macroscopical, description of giant gravitons
given in the literature.

\newpage
\sect{Introduction}

Recently, a non-Abelian action describing multiple Type IIA gravitational waves
has been derived in \cite{JL}, using Matrix string theory in a weakly curved
background. In this reference it was argued that this action represents an
appropriate description for a system of coincident Type IIA gravitons in the
strong coupling regime of the theory.

Several arguments supporting this claim were presented. One of them is that a 
T-duality transformation along the direction of propagation of the gravitons gives 
an action that can be identified as the non-Abelian action for multiple Type IIB 
fundamental strings \cite{BJL}. Moreover, this action for Type IIB F-strings is 
S-dual to the action for multiple D-strings, a fact that suggests that it is 
adequate to describe Type IIB strings in the strong coupling regime of the theory. 
Since T-duality does not change the regime of the coupling, the initial action for 
Type IIA gravitational waves should also be valid in the strong coupling regime.
A second argument is that, in the Abelian limit, the action of \cite{JL} is related 
to the perturbative action for massless particles by means of a world volume 
Legendre transformation, which typically relates the weak and strong coupling
regimes of a theory. Since the usual action for massless particles is obtained in 
the perturbative regime, the non-Abelian action of \cite{JL} is expected to be 
valid at strong coupling.

Since the strong coupling regime of Type IIA string theory is believed to be
M-theory, it is not difficult to extend the results of \cite{JL} to eleven 
dimensions. Basically, we just need to rewrite ten-dimensional background fields in 
terms of eleven-dimensional ones and rearrange the world volume fields in the 
ten-dimensional theory such that they are reinterpreted as the eleven-dimensional 
embedding scalars. It turns out that the action for M-theory gravitons derived in 
this manner is much simpler than the action describing Type IIA gravitons.
As in Type IIA, the Chern-Simons part contains the now familiar multipole
couplings that give rise to the dielectric effect \cite{Myers}. From them it is not
difficult to construct configurations of multiple coinciding gravitons
expanding into a fuzzy M2-brane. Furthermore, the Born-Infeld
action can easily be extended beyond the linear approximation used in the Matrix
theory calculation to a closed expression, by just demanding consistency with the
description of coincident D0-branes of \cite{Myers} after dimensional reduction.
This consistency provides a further check for this action.

Now, with a closed expression for the Born-Infeld action, valid beyond the linear 
approximation, an interesting possibility arises. It has been suggested before in
the literature (see for example \cite{DTV,Yolanda,BMN,JL}) that it should be 
possible to describe the giant gravitons of \cite{GST, GMT, HHI} in terms of
dielectric gravitational waves. More specifically, it is believed that in the limit 
where the number of gravitons becomes very large, this non-Abelian, microscopical 
description should match with the Abelian, macroscopical description of \cite{GST, 
GMT, HHI}, in terms of an expanding spherical brane with angular momentum, just as 
the spherical D2-brane with dissolved D0-charge of \cite{Emparan} is the large $N$ 
limit of the $N$ dielectric D0-branes expanding into a fuzzy D2 of \cite{Myers}. 
However, since a non-Abelian action for gravitational waves is only known up to 
linear order in the background fields and an $AdS_m \times S^n$ background cannot be
taken as a linear perturbation to Minkowski, it has not been possible, so far, to 
check whether dielectric gravitational waves really provide a microscopical
description for the giant gravitons in $AdS_m \times S^n$ of \cite{GST, GMT, HHI}.
With a closed expression for the Born-Infeld action for multiple waves it is now 
possible to construct explicitly configurations of multiple coinciding gravitons in
arbitrary backgrounds, which we can compare with the giant graviton
calculations.

The aim of this letter is precisely this. We will construct configurations of
$N$ eleven-dimensional gravitational waves polarising, by the dielectric
effect, into a fuzzy M2-brane in certain $AdS_m \x S^n$ backgrounds and compare
physical quantities, such as the energy of the configuration and the radius of the
expanded M2-brane, with the values arising from the Abelian calculation, as done
in \cite{GST, GMT}. If the latter is really the macroscopic description of the
former, one expects these physical quantities to agree in the large $N$ limit.
We will show that this is indeed the case.

The article is organised as follows. First, in Section 2, we briefly summarise the
construction from Matrix string theory of the non-Abelian action for Type IIA
gravitational waves of \cite{JL}. In Section 3 we present the non-Abelian action for
eleven-dimensional gravitons. Here we also show that the Born-Infeld part of the 
action can be extended beyond the linear approximation of the Matrix theory 
calculation. In Section 4 we discuss a toy model consisting of dielectric 
gravitational waves expanding in a fuzzy M2-brane in flat space and show the 
agreement between the microscopical and macroscopical descriptions. Finally, in 
Sections 5 and 6 we consider dielectric gravitational waves in more interesting
backgrounds and provide a non-Abelian description of giant gravitons in 
$AdS_7 \x S^4$ \cite{GST} and dual giant gravitons in $AdS_4 \x S^7$ \cite{GMT}.
We show the perfect agreement with the macroscopical descriptions of \cite{GST} and 
\cite{GMT} for large number of gravitons.

\sect{The action for multiple Type IIA gravitational waves}

In this section we recall the non-Abelian action for Type IIA gravitational waves
given in \cite{JL}. Since Matrix string theory describes strings with non-zero
light cone momentum, the same class of string states can, in the static gauge,
effectively be described in terms of massless particles. Indeed, it was shown in
\cite{JL} that Matrix string theory in a weakly curved background gives rise,
in the static gauge, to a non-Abelian generalisation of the action for a massless
particle, exhibiting the multipole terms first discovered in 
\cite{TvR2, TvR3, Myers} in the context of non-Abelian actions of D-branes.

The Matrix string theory action for weak background fields
\cite{Schiappa, BJL} is given by
\bea
S_{{\rm MST}} &=& \frac{1}{2\pi\beta^2} \int d\tau d\sigma
~\STr \Bigl\{
\tfrac{1}{2}
   \Bigl[ h_{ab} - \eta_{ab} (\phi - \tfrac{1}{2} h_{zz} )\Bigr] I_h^{ab}
\ + 2 B_{az} I_s^{az}
\ + \tfrac{1}{2} \Bigl[\phi + \tfrac{1}{2} h_{zz} \Bigr] I_h^{zz} \nn
&&
 - \tfrac{1}{2} \Bigl[ \phi
\ - \tfrac{3}{2} h_{zz}\Bigr] I_\phi
\ + C^{(1)}_a I_h^{az}
\ - C^{(3)}_{abz} I_s^{ab}
\ - C^{(1)}_z I_0^z
\ - h_{az} I_0^a
\ + 3 B_{ab} I_2^{abz}
\label{MST}
 \\ [.3cm]
&&
 +\ C^{(3)}_{abc} I_2^{abc}
\ + \tfrac{1}{12} C^{(5)}_{a_1 \ldots a_4 z} I_4^{a_1 \ldots a_4 z}
\ + \tfrac{1}{60} \tilde{B}_{a_1 \ldots a_5 z} I_4^{a_1 \ldots a_5}
\ + \tfrac{1}{48} N^{(7)}_{a_1 \ldots a_6 z} I_6^{a_1 \ldots a_6 z}
\nn
&& - \tfrac{1}{336} N^{(8)}_{a_1 \ldots a_7 z} I_6^{a_1 \ldots a_7}
\Bigr\},
\nonumber
\eea
where the indices $a = (0, i)$ and $i=1,...,8$. A special direction $Z$ comes
out specified in both the currents and the background fields. This direction is
the 9th direction of the 9-11 flip involved in the construction of the action.
Once we take static gauge, it is identified with the propagation direction of the
waves and appears as an isometry direction in the wave action.
The fields $C^{(2n-1)}$ denote the R-R $(2n-1)$-form potentials, $B$ the
Kalb-Ramond field and $\tB$ its Hodge dual, the NS-NS 6-form potential. The field
$N^{(7)}$ is the field that couples to the Type IIA Kaluza-Klein monopole,
and $N^{(8)}$ to the so-called $(6,1^2;3)$-brane\footnote{In the notation of
\cite{Hull}.}, an ``exotic'' brane (not predicted by the Type IIA spacetime 
supersymmetry algebra), first mentioned in \cite{BO, Hull, OPR}. The currents $I$ 
give the couplings between the Matrix (string) theory and the linearised background 
fields \cite{TvR2, TvR3, Schiappa} and are functions of the matrix valued transverse
coordinates $X^i$, the world volume scalar $A$ and their derivatives\footnote{See 
\cite{JL} for the interpretation of the field $A$.}. For their exact expression we 
refer to the appendix of \cite{JL}.

Filling in the expression for the currents and going to static gauge,
the linearised Chern-Simons action for multiple Type IIA gravitational waves, as
derived in \cite{JL}, is given by:\footnote{Up to rescalings. Reference \cite{JL}
used a special redefinition of the embedding scalars and world volume field
such that perturbative Type IIA string theory in the light-cone gauge was
recovered in the weak coupling limit. Here we are interested however in the
strong coupling regime.}
\begin{eqnarray}
\label{WACS}
S^{{\rm CS}}_{\rm{W_A}} &=& T_0 \int d\tau
~\STr \Bigl\{
- \  P [ \incl_k C^{(1)}] \wedge F
\ -  P [ \incl_k h]
\ + i   P [(\incl_X \incl_X)C^{(3)}]  \nn
&&
+\ i P[(\incl_X \incl_X) B] \wedge F
\ - \ i  P[\incl_{[A,X]} B]
\ +\ \tfrac12 P [(\incl_X \incl_X)^2 \incl_k \tB ]
\nn
&& 
-\ \tfrac12 P [(\incl_X \incl_X)^2 \incl_k C^{(5)} ] \wedge F
\ +\  P[(\incl_X \incl_X) \incl_{[A,X]} \incl_k C^{(5)} ]
\label{Wa} \\  [.3cm]
&& 
-\ \tfrac{i}{6} P [ (\incl_X \incl_X)^3 \incl_k N^{(7)}] \wedge F
+\ \tfrac{i}{2} P[(\incl_X \incl_X)^2
                                 \incl_{[A,X]} \incl_k N^{(7)} ]
\nn
&& 
- \ \tfrac{i}{6} P [ (\incl_X \incl_X)^3 i_k N^{(8)} ] \Bigr\}.
\nonumber
\end{eqnarray}
The contraction with $\incl_k$ denotes the interior product with the Killing vector
$k^\mu$ pointing along the direction of propagation $Z$ of the waves\footnote{The
identification of the isometric direction with the direction of propagation of the
waves comes from the analysis of the monopole term $P[\incl_k h]$, which is just the
linearised expression of the non-Abelian extension of the term giving momentum
charge to a single wave (see below).}, which can,
in adapted coordinates, be chosen as $k^\mu = \delta^\mu_z$. In the action
(\ref{WACS}), the direction of propagation appears as a special isometric
direction, in spite of which the whole action can be written in a manifestly
covariant way by making local the global gauge transformations along this direction
\begin{equation}
\delta X^\mu=\Lambda k^\mu,
\end{equation}
i.e. by working with a gauged sigma model. In (\ref{WACS}) the pull-backs into
the world volume are indeed defined in terms of gauge covariant
derivatives:\footnote{In a non-Abelian theory, the gauge covariant derivatives
are defined from $U(N)$-covariant derivatives
$D_\tau X^\mu= \part_\tau X^\mu + i [A_\tau, X^\mu]$. Here we are choosing however
the gauge where $A_\tau = 0$.}
\begin{equation}
\label{covder}
{\cal D}X^\mu = \part X^\mu-k^{-2}k_\rho \part X^\rho k^\mu,
\end{equation}
in such a way that the pull-backs associated
to the isome\-tric direction are cancelled out. Gauged sigma models of this type
have been used elsewhere, in order to describe the effective actions of
Kaluza-Klein monopoles \cite{BJO} and M-branes in massive eleven-dimensional
supergravity \cite{BLO}.

The inclusions $(\incl_X \incl_X)$ and $\incl_{[A,X]}$ are defined as
\bea
(\incl_X \incl_X \Sigma)_{\mu_1 ... \mu_p}
\equiv X^j X^i \Sigma_{ij\mu_1 ...\mu_p},
\hsp{2cm}
(\incl_{[A,X]} \Sigma)_{\mu_1 ... \mu_p}
\equiv [A,X^i]  \Sigma_{i\mu_1 ... \mu_p},
\eea
with $i=1,...,8$. The first type of contraction is the one giving rise to
the multipole couplings associated to the dielectric effect \cite{Myers}.
The world volume field $F$ in (\ref{WACS}) is defined
as $F= \part A$.

The Born-Infeld part of the action for Type IIA gravitational waves was not
given explicitly in \cite{JL}. One can however derive it from the Matrix theory
computation performed there and then taking static gauge. One
obtains, up to quartic order in the embedding scalars:
\begin{eqnarray}
\label{WABI}
S^{\rm BI}_{\rm W_A} &=& T_0\int d\tau ~{\rm STr}\Bigl\{
\frac12 h_{00}\Bigl( 1+\tfrac{1}{2}\dot{X}^2-\tfrac{1}{4}
[X,X]^2+\tfrac{1}{2}F^2-\tfrac{1}{2}[A,X]^2\Bigr)
\nn
&&
+ h_{0i}\dot{X}^i+\tfrac{1}{2} h_{ij}\Bigl( \dot{X}^i\dot{X}^j-
[X^i,X^k][X^k,X^j]+[A,X^i][A,X^j]\Bigr)
\nn
&&
+\half h_{zz}\Bigl( 1-\tfrac{1}{2}\dot{X}^2-\tfrac{1}{2}
F^2+\tfrac{1}{4}[X,X]^2+\tfrac{1}{2}[A,X]^2\Bigr)
\\ [.3cm]
&&
+\phi\Bigl( F^2-\tfrac{1}{2}[X,X]^2\Bigr)-
i  [A,X^i] \Bigl( B_{iz}+ B_{0z}\dot{X}^i\Bigr)
\nn
&&
+ C^{(1)}_0 F +  C^{(1)}_i \Bigl( \dot{X}^i F+
[X^i,X^j][A,X^j]\Bigr)
\nn
&&
+ i C^{(3)}_{0iz}\Bigl( [X^i,X^j]\dot{X}^j-
 [A,X^i]F\Bigr)-i\tfrac{1}{2}C^{(3)}_{ijz}[X^i,X^j]\Bigr\}.
\nonumber
\end{eqnarray}
This expression can be written in a fully covariant way as a gauged sigma model
using gauge covariant pull-backs into the world volume. We will postpone this
construction for the next section, where its relation to the
eleven-dimensional action will become clear.

It was also shown in \cite{JL} that the action there proposed to describe
multiple gravitational waves has the following expression as its
Abelian limit:\footnote{To be precise, the linearisation of this action,
since in the Matrix theory calculation one works to linear order in
the background fields.}
\begin{equation}
S_{\rm{W}} = \ -N  T_0 \int d \tau\  \Bigl\{ k^{-1}
                \sqrt{ | {\cal D} X^\mu {\cal D} X^\nu \ g_{\mu\nu} | }
                \ \  +  k^{-2} k_\mu \part X^\mu -\partial Z \Bigr\}.
\label{WAabelian}
\end{equation}
This action describes $N$ gravitational waves carrying momentum $T_0$ along the $Z$
direction. Therefore, the isometry direction is identified with the direction of
propagation of the waves.
As shown in \cite{JL}, (\ref{WAabelian}) is related by T-duality along $Z$
to the Nambu-Goto action for Abelian fundamental strings
with winding number along $Z$. Moreover, it can be mapped to
the action associated to $N$, Abelian, ten-dimensional massless particles through a
Legendre transformation that restores the dependence on the isometric direction
$Z$. Namely, (\ref{WAabelian}) can be related to the following
action, in terms of an auxiliary metric $\gamma$ on the world line
(see \cite{JL}):
\begin{equation}
\label{masspar}
S[\gamma]=-\frac{NT_0}{2}\int d\tau\ \sqrt{|\gamma|}\ \gamma^{-1}
\partial X^\mu \partial X^\nu g_{\mu\nu} ,
\end{equation}
where $\mu=0,\dots,9$. Therefore,
the Matrix string calculation describes Type IIA waves by means of
a non-Abelian extension of the particular description of massless particles
with non-zero momentum given by (\ref{WAabelian}).
Since (\ref{masspar}) is a perturbative action, (\ref{WAabelian})
is expected to be valid in the strong coupling regime, given that world volume
duality transformations typically relate the weak and strong coupling
regimes of a theory.

\sect{The action for multiple gravitational waves in eleven dimensions}

We now want to  uplift the non-Abelian action for Type IIA gravitational waves
to eleven dimensions.  For this we use the following linearised relations between 
eleven- (denoted with hats) and ten-dimensional (unhatted) background fields:
\be
\begin{array}{lll}
{\hat h}_{\mu\nu} = h_{\mu\nu}-2\phi/3\ \eta_{\mu\nu}, \hsp{.2cm}&
{\hat h}_{\mu 11} = C^{(1)}_\mu, &
{\hat h}_{11,11}=4\phi/3, \\ [.3cm]
{\hat C}_{\mu_1\mu_2\mu_3} = C^{(3)}_{\mu_1\mu_2\mu_3}, &
{\hat C}_{\mu_1\mu_2 11} = B_{\mu_1\mu_2}, &
(\incl_{\hat k}{\hat N}^{(8)})_{\mu_1\dots\mu_7} =
(\incl_k N^{(8)})_{\mu_1\dots\mu_7},\\ [.3cm]
{\hat {\tilde C}}_{\mu_1\dots\mu_6} = {\tilde B}_{\mu_1\dots\mu_6},&
{\hat {\tilde C}}_{\mu_1\dots\mu_5 11} = -C^{(5)}_{\mu_1\dots\mu_5}, \hsp{.2cm} &
(\incl_{\hat k}{\hat N}^{(8)})_{\mu_1\dots\mu_6 11} =
(\incl_k N^{(7)})_{\mu_1\dots\mu_6},
\end{array}
\ee
where $(\incl_k {\hat N}^{(8)})$ is the field that couples minimally to the
Kaluza-Klein monopole of M-theory and $N^{(7)}$ and $N^{(8)}$ are the Type IIA
gravitational background fields in (\ref{WACS}).

Regarding the eleven-dimensional embedding coordinates, they can be
straightforwardly obtained from the ten-dimensional world volume fields, if we
identify the world volume scalar $A$ as the eleventh embedding coordinate:
\begin{equation}
\hX^i = X^i, \hsp{1.5cm}
\hX^{9}= A, \hsp{1.5cm}
\hat{Z} = Z.
\label{X=A}
\end{equation}
With these identifications, it is not difficult to see that the currents in
(\ref{MST}) can be written in eleven dimensions in the static gauge as
$(\hi=1,...,9)$
\be
\begin{array}{ll}
\hI_\phi = \unity - \half  (\dhX)^2
             - \tfrac{1}{4} [\hX, \hX]^2, \hsp{1cm} &
\hI_h^{\hi\hj} =  \dhX{}^\hi \dhX{}^\hj
               -  [\hX^\hi, \hX^\hk][\hX^\hk, \hX^\hj],
\\[.3cm]
\hI_h^{00} = \unity + \half (\dhX)^2
             - \tfrac{1}{4} [\hX, \hX]^2, &
\hI_s^{0\hi} = - \tfrac{i}{2} [\hX^\hi, \hX^\hj] \dhX{}^\hj,
\\[.3cm]
\hI_h^{0\hi} = \dhX{}^\hi, &
\hI_s^{\hi\hj} = \tfrac{i}{2} [\hX^\hi, \hX^\hj],
\end{array}
\ee
and
\be
\begin{array}{ll}
\hI_0^0 = \unity,&
\hI_4^{0\hi\hj\hk\hl} = -\tfrac{3}{2}
                    [\hX^\hi, \hX^{[\hj}][\hX^\hk, \hX^{\hl]}],
\\ [.3cm]
\hI_0^\hi =  \dhX{}^\hi,&
\hI_4^{\hi\hj\hk\hl\hm} = -\tfrac{15}{2}
               \dhX{}^{[\hi}[\hX^\hj, \hX^\hk][\hX^\hl, \hX^{\hm]}],\\ [.3cm]
\hI_2^{0\hi\hj} = -\tfrac{i}{6} [\hX^\hi, \hX^\hj],&
\hI_6^{0\hi\hj\hk\hl\hm\hn} = i
               [\hX^{[\hi}, \hX^\hj][\hX^\hk, \hX^\hl][\hX^\hm, \hX^{\hn]}],
                                                                   \\ [.3cm]
\hI_2^{\hi\hj\hk} = -\tfrac{i}{2}\dhX{}^{[\hi} [\hX^\hj, \hX^{\hk]}],
 \hsp{1cm}&
\hI_6^{\hi\hj\hk\hl\hm\hn\hp} = 7i
                  \dhX{}^{[\hi}[\hX^\hj, \hX^\hk][\hX^\hl, \hX^\hm]
                           [\hX^\hn, \hX^{\hp]}].
\end{array}
\label{CScurrents}
\ee
The linear action (\ref{MST}) can then be written in the static gauge as
\bea
S &=&  {\hat T}_0 \int d\tau~ {\rm STr}\Bigl\{
\half \Bigl[ \hh_{\ha\hb} + \half \hh_{zz}\eta_{\ha\hb} \Bigr] \hI_h^{\ha\hb}
\ + \ \tfrac{3}{4} \hh_{zz} \hI_\phi
\ - \ \hC_{\ha\hb z} \hI_s^{\ha\hb}
\ -  \ \hh_{\ha z} \hI_0^\ha  \nn
&&  \hsp{2cm}
+\ \hC_{\ha\hb\hc} \hI_2^{\ha\hb\hc}
\ + \ \tfrac{1}{60} {\hat{\tilde C}}_{\ha_1...\ha_5 z} \hI_4^{\ha_1...\ha_5}
\ - \ \tfrac{1}{336} {\hat N}^{(8)}_{\ha_1...\ha_7 z} \hI_6^{\ha_1...\ha_7}
\Bigr\},
\label{lin11}
\eea
and we find the following expression for the linearised Chern-Simons action
associated to coinciding M-theory gravitons:
\bea
\label{MwavesCS}
S^{{\rm CS}}_{{\hat {\rm W}}} &=& {\hat T}_0 \int d\tau~ {\rm STr}\Bigl\{
- P[\incl_{\hat k}{\hat h}]
+ i  P[(\incl_{\hat X}\incl_{\hat X}){\hat C}]
+ \tfrac{1}{2} P[ (\incl_{\hat X}\incl_{\hat X})^2
                             \incl_{\hat k} {\hat {\tilde C}} ] \\ [.3cm]
&& \hsp{5cm}
- \tfrac{i}{6} P[ (\incl_{\hat X}\incl_{\hat X})^3
                               \incl_{\hat k} {\hat N}^{(8)} ] \Bigr\}.
\nonumber
\eea
This action consists of a monopole term $P[\incl_{\hat k}{\hat h}]$ plus a series
of multipole, Myers-like couplings. Like the action for non-Abelian Type IIA 
gravitational waves, it contains the direction of propagation of the waves as a 
special isometric direction, although the whole action can still be written in a 
manifestly covariant way by making local the global scale transformations along this
direction. The gauge covariant derivatives implicit in the pull-backs of this action
are defined as in (\ref{covder}), but now in terms of eleven-dimensional fields.

The monopole term  $P[\incl_{\hat k}{\hat h}]$ is similar to the monopole term
encountered in the action of ten-dimensional gravitational waves \cite{JL},
indicating that we are dealing with eleven-dimensional gravitational waves with 
momentum in the $\hat{Z}$ direction. The multipole couplings in (\ref{MwavesCS}) 
are typical non-Abelian, Myers-like, couplings. In particular, the term
$P[(\incl_{\hat X}\incl_{\hat X}){\hat C}]$ corresponds to the dielectric term
for gravitational waves introduced in \cite{BMN} in the Matrix model of DLCQ 
eleven-dimensional pp-waves, in order to make the action
supersymmetric. The multipole terms of (\ref{MwavesCS}) were also predicted
previously in \cite{Yolanda} via string duality arguments.

In a similar way, the Born-Infeld part of the action for M-theory gravitons is
derived from (\ref{lin11}) or, equivalently, by uplifting the ten-dimensional
action (\ref{WABI}). We obtain:
\begin{eqnarray}
\label{MwavesBIl}
S^{\rm BI}_{{\hat {\rm W}}} &=& -{\hat T}_0\int d\tau~ {\rm STr} \Bigl\{
   \half {\hat h}_{00}\Bigl(\unity+\half \dhX{}^2-\tfrac{1}{4} [\hX,\hX]^2 \Bigr)
   + \ {\hat h}_{0\hi}\dhX{}^{\hi}
 \nn
&& +\ \half {\hat h}_{\hi\hj} \Bigl(\dhX{}^{\hi}\dhX{}^{\hj}
                 - [\hX^{\hi},\hX^{\hk}][\hX^{\hk},\hX^{\hj}] \Bigr)
   +\ \half {\hat h}_{zz} \Bigl( \unity  -\half \dhX{}^2
                 +\tfrac{1}{4} [\hX,\hX]^2 \Bigr) \nn
&&+\ i  {\hat C}_{z0\hi}[\hX^\hi,\hX^\hj]\dhX{}^{\hj}
  - \ \tfrac{i}{2}{\hat C}_{z\hi\hj}[\hX^\hi,\hX^\hj] \ \Bigr\},
\end{eqnarray}
up to quartic order in the embedding scalars. In fact, it is not too complicated to show
that this action is the linear expansion of the following BI action:
\begin{equation}
\label{MwavesBI}
S^{\rm BI}_{{\hat {\rm W}}}={\hat T}_0\int d\tau ~ {\rm Str}\Bigl\{
{\hat k}^{-1} \sqrt{
   - P[{\hat E}_{00}
   + {\hat E}_{0{\hi}}({\hat Q}^{-1}-\delta)^{\hi}_{\hk}
          {\hat E}^{{\hk}{\hj}}{\hat E}_{{\hj}0}]
\ {\det}({\hat Q}^{\hi}_{\hj})}
\Bigr\}.
\end{equation}
Here we take
\begin{equation}
\label{Ehat}
{\hat E}_{{\hat \mu}{\hat \nu}} \equiv
{\hat {\cal G}}_{\hmu\hnu}
+ {\hat k}^{-1} (\incl_{\hat k} {\hat C})_{{\hat \mu}{\hat \nu}},
\end{equation}
while ${\hat E}^{\hi\hj}$ denotes the inverse of ${\hat E}_{\hi\hj}$, and
\begin{equation}
{\hat Q}^\hi_\hj \equiv
\delta^\hi_\hj + i {\hat k} [\hX^\hi,\hX^\hk] {\hat E}_{\hk\hj}\ .
\end{equation}
Note that the action (\ref{MwavesBI}) is also a gauged sigma model, as expected
for the action for gravitational waves. The isometry direction ${\hat Z}$ is
projected out through the effective metric
${\hat {\cal G}}_{\hmu\hnu} = \hg_{\hmu\hnu} - \hk^{-2} \hk_\hmu \hk_\hnu$
of (\ref{Ehat})
and the contraction of the 3-form with the Killing direction,
$(\incl_{\hat k} {\hat C})$.

Moreover, (\ref{MwavesBI}) reduces to the Born-Infeld action for coincident
D0-branes, derived in \cite{Tseytlin, Myers}, when reduced along the direction of
propagation of the waves, $\hat{Z}$. Similarly, the non-Abelian Chern-Simons action
for multiple D0-branes \cite{Myers, TvR3} is recovered, now to linear order in the
background fields, when the Chern-Simons action (\ref{MwavesCS}) is reduced over
the isometry direction. We believe this is a non-trivial check that confirms our
interpretation of the actions (\ref{MwavesCS}), (\ref{MwavesBIl}) and
(\ref{MwavesBI}) as describing M-theory gravitons.

One further check of $S_{\rm {\hat W}}=S^{\rm BI}_{\rm {\hat W}} +
S^{\rm CS}_{\rm {\hat W}}$ is that, in the Abelian limit, it describes
eleven-dimensional massless particles. In this limit
the full action given  by (\ref{MwavesCS}) and (\ref{MwavesBIl}) becomes the
linearised expansion of the action:
\begin{equation}
\label{MWabelian}
S_{{\hat {\rm W}}}=-N{\hat T}_0 \int d\tau \Bigl\{ {\hat k}^{-1}
\sqrt{|\partial \hX^{\hat \mu} \partial \hX^{\hat \nu}
          (\hg_{{\hat \mu}{\hat \nu}} - \hk^{-2}\hk_{\hat \mu} \hk_{\hat \nu})|}
      + \hk^{-2}\hk_\hmu \part \hX^\hmu -\partial \hZ\Bigr\} ,
\end{equation}
which, as we showed in the analogous Type IIA case \cite{JL}, describes
eleven-dimensional massless particles carrying momentum along $\hat{Z}$. This 
action can be mapped to the usual perturbative action for massless particles by
means of a Legendre transformation, which restores the dependence on the
direction of propagation.

Finally, we can use the full Born-Infeld action for M-theory gravitons given by
(\ref{MwavesBI}) to construct a Born-Infeld action for Type IIA gravitational
waves valid beyond the linear approximation. This calculation is
complicated in the general case, so we have taken a truncation of the
backgrounds: $C^{(1)}=B=0$, plus a vanishing world volume field $A$. This action
contains still non-Abelian couplings of interest for the study of
dielectric configurations. We have obtained
\begin{equation}
\label{IIAwavesBIc}
S^{\rm BI}_{{\rm W}_A}=-T_0\int d\tau~ {\rm Str}\Bigl\{ k^{-1}
\sqrt{-P[E_{00}+E_{0i}(Q^{-1}-\delta)^i_k
E^{kj}E_{j0}] \ {\rm det} (Q^i_j )}
\ \Bigr\},
\end{equation}
where
\begin{eqnarray}
E_{\mu\nu} &=& g_{\mu\nu}-k^{-2}k_\mu k_\nu
                   + k^{-1}e^\phi (\incl_k C^{(3)})_{\mu\nu},  \\
Q^i_j &=& \delta^i_j + i [X^i,X^k]e^{-\phi}k E_{kj}.
\end{eqnarray}
In the next sections, we will use the action given by (\ref{MwavesBI}) and
(\ref{MwavesCS}) 
to construct non-Abelian giant graviton configurations and relate these to some of 
the Abelian giant gravitons known in the literature. 
 
\sect{Giant gravitons in flat space}
\label{giantgrav}

\subsection{The non-Abelian description}

The explicit form for the action describing multiple coincident gravitons in
M-theory allows the study of giant graviton configurations from the microscopical
point of view of the expanding gravitons, directly in M-theory. From now on we will 
restrict to eleven-dimensional spacetimes, so we will omit the hats.
The action given by (\ref{MwavesCS}) and (\ref{MwavesBI}) describes M-theory 
gravitons propagating along the isometric direction $Z$. The BI action 
(\ref{MwavesBI}) contains in particular a non-Abelian coupling to the 3-form 
potential\footnote{See the last terms of (\ref{MwavesBIl}).}, which can be written 
in the form of a magnetic moment coupling, similar to those studied in \cite{Myers, 
DTV}. The potential for a static configuration of M-theory gravitons subject to an 
external $F^{(4)}_{zijk}$ field strength is given by
\begin{equation}
\label{Mpot}
V(X)=T_0 \ {\rm STr}\Bigl\{
           \unity-\frac14 [X,X]^2
        + \frac{i}{3} X^j X^i X^k F^{(4)}_{zijk}\Bigr\}
\end{equation}
to leading order for the Born-Infeld potential.
This potential has a stable solution of the form:
\begin{equation}
\label{Msol}
F^{(4)}_{zijk}=f \epsilon_{ijk} \ , \qquad
X^i=-\frac{f}{4} J^{i},
\end{equation}
with $J^{i}$ forming an $N\times N$ representation of $SU(2)$, which we interpret
in terms of $N$ gravitons expanding into an M2-brane transverse to their
direction of propagation. The M2-brane has the topology of a fuzzy sphere with
radius
\begin{equation}
\label{Rmic}
R^2=\frac{f^2}{16}(N^2-1),
\end{equation}
and energy

\begin{equation}
\label{Emic}
E=T_0\Bigl\{N-\frac{1}{384}f^4 N(N^2-1)\Bigr\}.
\end{equation}
This solution is identical to the dielectric solution of \cite{Myers},
the only difference being that our configuration carries a magnetic moment
instead of an electric one with respect to the external field.

Consider now the case in which the gravitons propagate along a compact direction, 
i.e.~$Z=X^{9}$. Reducing (\ref{Mpot}) and (\ref{Msol}) along the compact direction 
we can interpret the potential above as describing static Type IIA D0-branes subject
to an external magnetic field $H_{ijk}$. The stable solution given by (\ref{Msol}) 
is then interpreted in terms of D0-branes expanded into a fuzzy D2-brane carrying 
magnetic moment. This solution in Type IIA was studied by Myers in \cite{Myers} and 
then used in \cite{DTV} in order to describe, in the Type IIA limit, the expansion 
of M-theory gravitons, propagating along the eleventh direction, into a transverse 
fuzzy M2-brane. Our description for eleven-dimensional gravitons allows to study 
this configuration directly in M-theory and, more importantly, for an arbitrary 
direction of propagation of the gravitons.

When the gravitons propagate in a direction different from the direction we 
compactify on, the description in the Type IIA limit is in terms of Type IIA
gravitational waves propagating along the same direction. This happens because
eleven-dimensional waves give rise to ten-dimensional waves when the reduction
from eleven to ten dimensions takes place in a direction transverse to the
direction of propagation. It is straightforward to see that the reduction of the
potential (\ref{Mpot}) and of the solution (\ref{Msol}) gives rise to Type IIA
waves in an external $F^{(4)}_{zijk}$ magnetic field, expanding into a fuzzy
D2-brane transverse to their direction of propagation \cite{JL}.

$N$ D0-branes with non-zero velocity have been used as well to describe M-theory
gravitons in the Type IIA limit \cite{DTV}. Given that D0-branes are equivalent
to M-theory gravitons propagating along the eleventh direction, one describes in
this manner M-theory gravitons moving both along the eleventh direction and along
the direction of propagation of the D0-branes. Our M-theory analysis allows
for a description of expanding gravitons with no velocity along the eleventh
direction.

\subsection{The Abelian description}

Let us now analyse the M-theory solution that we have just discussed from the
point of view of its dual formulation in terms of the M2-brane action. We should
find a solution corresponding to an M2-brane wrapping a classical sphere with the
same quantum numbers of the gravitons. As in \cite{Myers}, the two
descriptions should agree in the large $N$ limit. The Ansatz for the
4-form potential of (\ref{Msol}) reads, in spherical coordinates:
\begin{equation}
F^{(4)}_{zr\theta\phi}=f r^2 \sin{\theta}
\end{equation}
which implies a 3-form potential:
\begin{equation}
\label{hatCM2}
C^{(3)}_{z\theta\phi}=-\frac{1}{3}f r^3\sin{\theta}.
\end{equation}
The effective action associated to a spherical M2-brane with world volume
directions $(\tau,\theta,\phi)$ and radius $R$ subject to the external potential
(\ref{hatCM2}) and with velocity along an arbitrary $Z$-direction is given by:
\begin{equation}
\label{M2spher}
S_{\rm M2}= -4\pi T_2\int d\tau \Bigr[R^2\sqrt{1-\dot{Z}{}^2}
                  -\tfrac{1}{3} f R^3 \dot{Z} \Bigr] ,
\end{equation}
where the angular world volume coordinates have been integrated out.
The velocity of the M2-brane has to be computed from the condition that it must
have $N$ graviton charge dissolved in its world volume or, in other words, that
it should carry $N$ units of linear momentum. This implies that
\begin{equation}
P_{z}=\frac{N}{2}\sin{\theta},
\end{equation}
such that when integrating over $\theta$ and $\phi$ we get the factor of $4\pi$
necessary to match the tension of the M2-brane with the tension of a single
graviton, i.e.~$T_0=2\pi T_2$. This value for the momentum can be seen arising
more explicitly in the case in which the M2-brane propagates along the eleventh
direction. In this case we can make contact with the description in the Type IIA
limit in terms of a spherical D2-brane with $N$ D0-branes dissolved in its world
volume. In the Type IIA limit the coupling
\begin{equation}
\int_{S^2} C^{(1)}\wedge F,
\end{equation}
in the world volume effective action of a single D2-brane, implies that we must
take
a Born-Infeld magnetic flux
\begin{equation}
F_{\theta\phi}=\frac{N}{2}\sin{\theta}
\end{equation}
in order to have $N$ $C^{(1)}$-charge, and therefore $N$ D0-branes, dissolved in
the world volume. This magnetic flux is reinterpreted in eleven dimensions as the
momentum carried by the M2-brane. In order to see this recall that
the D2-brane and the M2-brane effective actions
are related by means of a world volume duality transformation that maps the
Born-Infeld field of the D2-brane into the extra eleventh scalar of the M2-brane.
This transformation has a simple formulation in terms of a canonical
transformation in the world volume phase space \cite{yo}. In fact, given that we
are dealing with non-static configurations it is more useful to discuss the
dynamics in terms of Hamiltonians. As shown in \cite{yo}, the canonical
transformation that relates the D2 and M2-brane effective actions simply works at
the level of the Hamiltonians transforming the Born-Infeld magnetic field of the
D2-brane into the canonical momentum associated to the eleventh direction of the
M2-brane, and the canonical momentum associated to the Born-Infeld field into
the spatial derivative of the eleventh direction:
\begin{equation}
P_{9}= \frac12 \epsilon^{0\alpha\beta}F_{\alpha\beta}, \hsp{2cm}
\Pi^\alpha=-\epsilon^{0\alpha\beta}\partial_{\beta} X^{9} .
\end{equation}
For our spherical M2-brane we then have that
$P_{9}=F_{\theta\phi}=\frac{N}{2}\sin{\theta}$. In general, for an arbitrary
direction of propagation $Z$, we have that $P_z=\frac{N}{2}\sin{\theta}$,
which implies a value for the velocity:
\begin{equation}
\dot{Z}=\frac{\frac{1}{2}N-\frac{1}{3}f R^3}
                  {\sqrt{R^4+(\frac{1}{2}N-\frac{1}{3}fR^3)^2}},
\end{equation}
as derived from (\ref{M2spher}). We then have that the Hamiltonian for the
spherical M2-brane reads:
\begin{equation}
\label{Hammac}
H=4\pi T_2 \ \sqrt{R^4+(\frac{N}{2}-\frac{f}{3}R^3)^2},
\end{equation}
where we have integrated over $\theta$ and $\phi$. This Hamiltonian can be
approximated in the large $N$ limit by:
\begin{equation}
H=NT_0 \ \Bigl( 1-\frac23 \frac{f}{N}R^3 + \frac{2}{N^2}R^4\Bigr),
\end{equation}
for which we find a stable minimum for
\begin{equation}
\label{RlargeN}
R=\frac{Nf}{4} ,
\end{equation}
with energy
\begin{equation}
\label{ElargeN}
E=N T_0\Bigl(1-\frac{f^4 N^2}{384}\Bigr) .
\end{equation}
The values of (\ref{RlargeN}) and (\ref{ElargeN}) are just the large $N$ limit
expressions of the radius and energy given by (\ref{Rmic}) and (\ref{Emic}).
Therefore, we can conclude that the microscopical and macroscopical descriptions
do indeed agree up to $1/N$ corrections.

\vspace{0.2cm}

In the next sections we consider more interesting configurations. We will
describe microscopically, in terms of expanding gravitational waves in
M-theory, the giant
graviton of \cite{GST} in an $AdS_7 \times S^4$ background,
as well as the so-called dual giant graviton of
\cite{GMT} in an $AdS_4 \times S^7$ spacetime. In these two cases the gravitons
expand into a non-commutative 2-sphere living in the spherical and
anti-de-Sitter part of the geometry respectively. We show that in the limit in
which the number of gravitons goes to infinity there is perfect agreement between
our description and the macroscopical description in terms of a classical
spherical M2-brane of \cite{GST} and \cite{GMT}.

\sect{Giant gravitons in $AdS_7 \times S^4$}

\subsection{The non-Abelian calculation}

Following \cite{GST} we would like to find a stable solution where $N$ gravitons
have expanded into a fuzzy M2-brane on the spherical part of the geometry.

The full line element for the metric on $AdS_7 \times S^4$ has the form
$ds^2=ds_{AdS}^2 + ds_{S}^2$, where

\begin{equation}
\label{AdS7}
ds_{AdS}^2=-(1+\frac{r^2}{{\tilde L}^2})dt^2
           +  \frac{dr^2}{1+\frac{r^2}{{\tilde L}^2}}
           + r^2 d\Omega_{5}^2,
\end{equation}
and
\begin{equation}
\label{S4}
ds_{S}^2=L^2 (d\theta^2 + {\cos}^2{\theta}d\phi^2
                        + {\sin}^2{\theta} \ d\Omega_2^2).
\end{equation}
The radius of curvature of $AdS_7$ is ${\tilde L}$ while that for $S^4$ is
$L$, and they are related as $L={\tilde L}/2$. We are looking for a solution in
which the gravitons in this metric expand into a fuzzy $S^2$ of radius
$L\sin{\theta}$, i.e.~the fuzzy version of a classical 2-sphere of this
radius, contained in (\ref{S4}). Let us choose coordinates in this sphere as:
\begin{equation}
d\Omega_{2}^2=d\chi_1^2+{\sin}^2{\chi_1} \ d\chi_2^2.
\end{equation}
As in \cite{GST} we look for a configuration: $\theta={\rm const}$,
$\phi=\phi(\tau)$, $r=0$, where we have taken static gauge and identified
$t=\tau$. This implies that the fuzzy $S^2$ has a constant radius. Our Ansatz for
the fuzzy $S^2$ is then
\begin{equation}
X^i=\frac{L\sin{\theta}}{\sqrt{N^2-1}}J^i,
\end{equation}
where $J^i,~(i=1,2,3)$ form an $N\times N$ representation of $SU(2)$.
With this Ansatz
\be
(X^1)^2+(X^2)^2+(X^3)^2=L^2{\sin}^2{\theta} \ \unity.
\ee The rest of the coordinates remain Abelian, i.e.~proportional to $\unity$.
On the other hand, the 3-form potential on $S^4$, given by
\begin{equation}
C^{(3)}_{\phi\chi_1\chi_2}=L^3 {\sin}^3{\theta} \sin{\chi_1},
\end{equation}
reads in the Cartesian coordinates $(X^1,X^2,X^3)$ above:
\begin{equation}
C^{(3)}_{\phi ij}=-\epsilon_{ijk} X^k.
\end{equation}

We now have all the elements to try to find a stable solution. The
effective action describing $N$ gravitons with velocity along $\phi$ is
described by (\ref{MwavesBI}), with $\phi$ playing the role of the
isometric direction in this action. With the particular background described
above there is no contribution of the Chern-Simons action, given that the
coupling to the 3-form potential in (\ref{MwavesCS}) is computed with
gauge covariant derivatives, and these eliminate the contribution of the
$\phi$ component. The 3-form potential couples however in the Born-Infeld
part of the action, through
$E_{ij}={\cal G}_{ij}+k^{-1}C^{(3)}_{\phi ij}$.

Substituting the Ansatz above into the world volume action (\ref{MwavesBI}) 
we obtain the following 
potential:\footnote{Recall that in our description of the gravitons the direction of
propagation occurs as an isometry direction, and therefore we are dealing with a 
static configuration, for which we can compute the potential as minus the 
Lagrangian.}
\begin{equation}
\label{potexact}
V(X)= \frac{T_0}{L\cos{\theta}}~{\rm STr}\Bigl\{ 
\sqrt{\unity - \frac{4L\sin{\theta}}{\sqrt{N^2-1}}X^2 
             + \frac{4L^4{\sin}^2{\theta}\cos^2 \theta }{N^2-1} X^2
             + \frac{4L^2{\sin}^2{\theta}}{N^2-1}X^2X^2} \ \Bigr\},
\end{equation}
where we have taken into account that some contributions to $\det(Q^i_j)$
do in fact vanish after taking the symmetrised average involved in the symmetrised
trace prescription in (\ref{MwavesBI}). Using the definition of the symmetrised
trace we can now proceed and compute the potential.

Since we are interested in the comparison with the Abelian calculation of
\cite{GST}, it is convenient to look at the large $N$ limit\footnote{See the
Conclusions for the finite $N$ calculation to leading order in the Born-Infeld
potential, i.e. the usual approximation, taken for instance
in the non-Abelian calculation in \cite{Myers}.}.
When $N$ is large 
\be
\begin{array}{ll}
{\rm STr}\{(X^2)^n\}& \approx {\rm Tr}\{(X^2)^n\} + {\cal O} (\tfrac{1}{N^2-1})
\\[.3cm]
& = L^{2n}{\sin}^{2n}{\theta} N + {\cal O} (\tfrac{1}{N^2-1}),
\end{array}
\label{STr-Tr}
\ee
since the commutators involved in the rewriting of (\ref{STr-Tr}) can be neglected.
Therefore, the potential (\ref{potexact}) can be written as:
\begin{equation}
V(\theta) = \frac{NT_0}{L\cos{\theta}}
       \sqrt{ 1 - \frac{4L^3}{\sqrt{N^2-1}}{\sin}^3{\theta}
                + \frac{4L^6}{N^2-1}{\sin}^4{\theta}},
\end{equation}
or, introducing the $\cos{\theta}$ inside the square root:
\begin{equation}
\label{potmicr1}
V(\theta) = \frac{NT_0}{L}
         \sqrt{ 1 + \tan^2{\theta}
                 \Bigl(1-\frac{2L^3}{\sqrt{N^2-1}}\sin{\theta} \Bigr)^2},
\end{equation}
where we are neglecting terms of order $(N^2-1)^{-\frac{5}{2}}$ in the
expansion of (\ref{potexact}).
This potential admits two minima, at $\sin{\theta}=0$, the point-like
graviton, and at
\begin{equation}
\sin{\theta}=\frac{\sqrt{N^2-1}}{2L^3}.
\label{giantgr}
\end{equation}
Both these minima have an energy
\begin{equation}
E=\frac{NT_0}{L}=\frac{P_\phi}{L},
\end{equation}
associated to a massless particle with angular momentum $P_\phi$. Moreover, the
angular momentum has an upper bound derived from the fact that
\begin{equation}
\label{solmicr1}
\sin{\theta}=\frac{\sqrt{N^2-1}}{2L^3}\leq 1,
\end{equation}
which implies
\begin{equation}
\label{uppermicr}
P_\phi\leq T_0 \sqrt{4L^6 + 1}.
\end{equation}
The Hamiltonian, the solution and the upper bound correspond, in the large $N$ 
limit, to the same physical quantities found in \cite{GST}, as we will show below 
when we discuss the Abelian calculation.

\subsection{The Abelian calculation}

The macroscopical description of the giant graviton solution in an
$AdS_7\times S^4$ background was the
calculation presented by \cite{GST}, in terms of a classical M2-brane
carrying momentum and expanding into the spherical part of the background
geometry. We briefly summarise these results, following closely \cite{GMT}, in
order to see the explicit agreement with our microscopical calculation.

In this case we have a test brane solution consisting on an
M2-brane\footnote{In our conventions this is in fact an anti M2-brane.} whose
world volume lays on the classical sphere with radius $L\sin{\theta}$.
The Hamiltonian was computed in \cite{GST} and found to be:
\begin{equation}
\label{potmacr1}
H=\frac{P_{\phi}}{L}\sqrt{1+{\tan}^2{\theta}\Bigl( 1-{\frac{{\tilde
N}}{P_\phi}\sin{\theta}\Bigr)^2}},
\end{equation}
where ${\tilde N}$ is an integer that emerges through the quantisation
condition of the 3-form flux on $S^4$
\begin{equation}
\label{quanti}
4\pi T_2 = \frac{\tilde N}{L^3},
\end{equation}
$T_2$ being the tension of the M2-brane.

The Hamiltonian (\ref{potmacr1}) has two stable minima, one for $\sin{\theta}=0$
and another for $\sin{\theta}=P_\phi/{\tilde N}$. The value of the Hamiltonian
corresponds in both cases to $E=P_\phi/{L}$, i.e.~to a massless particle. The
condition that $\sin{\theta}\leq 1$ implies an upper bound for the momentum
$P_\phi\leq {\tilde N}$.

In terms of the expanding gravitons, we have that the momentum $P_\phi=N T_0$.
Moreover, the quantisation condition (\ref{quanti}) reads, in terms of the
tension of the gravitons:
\bea
\label{NtildeN}
T_0=\frac{{\tilde N}}{2L^3}.
\eea
Therefore we can rewrite the non-trivial solution and the Hamiltonian as
\bea
\sin{\theta}&=&\frac{P_\phi}{{\tilde N}}=\frac{N}{2L^3},\nn
H&=&\frac{NT_0}{L}\sqrt{1+{\tan}^2{\theta}\Bigl(
1-{\frac{2L^3}{N}\sin{\theta}\Bigr)^2}},
\eea
which coincide with the first orders in a $1/N$ expansion of the microscopical 
solution (\ref{giantgr}) and the Hamiltonian (\ref{potmicr1})!

The same agreement is found regarding the upper bound for the angular momentum,
given that using (\ref{NtildeN}):
\begin{equation}
P_\phi\leq {\tilde N}=2L^3 T_0
\end{equation} 
which agrees with the large $N$ limit of (\ref{uppermicr}). 

\sect{Dual Giant Gravitons in $AdS_4\times S^7$}

\subsection{The non-Abelian calculation}

Following \cite{GMT} we would like to find a stable solution where $N$
gravitons have expanded into a fuzzy $S^2$ that now lives in the $AdS$ part of
the geometry. These configurations were referred to as dual giant gravitons by
the authors of \cite{GMT}.

The full line element for the metric on $AdS_4\times S^7$ takes the form
$ds^2=ds_{AdS}^2+ds_{S}^2$, where
\begin{equation}
ds_{AdS}^2=-(1+\frac{r^2}{{\tilde L}^2})dt^2
            + \frac{dr^2}{(1+\frac{r^2}{{\tilde L}^2})}
              + r^2 d\Omega_2^2
\end{equation}
and
\begin{equation}
ds_{S}^2=L^2(d\theta^2+{\cos}^2{\theta}d\phi^2+{\sin}^2{\theta} \ d\Omega_5^2) .
\end{equation}
The radius of curvature ${\tilde L}$ of $AdS_4$ and $L$ of $S^7$ are now
related as $L=2{\tilde L}$. In this case we are interested in a solution for
which the gravitons expand into a fuzzy 2-sphere with radius $r$ contained in
the $AdS$ part of the geometry, since this is the non-commutative version of the
solution in \cite{GMT}. We will choose coordinates in this sphere as:
\begin{equation}
d\Omega_2^2=d\alpha_1^2+{\sin}^2{\alpha_1} \ d\alpha_2^2.
\end{equation}
Our $SU(2)$ Ansatz for the fuzzy $S^2$ is now:
\begin{equation}
X^i=\frac{r}{\sqrt{N^2-1}}J^i,
\end{equation}
for $i,j=1,2,3$. The 3-form potential on the $AdS$ part of the geometry reads:
\begin{equation}
C^{(3)}_{0\alpha_1\alpha_2}=-\frac{r^3}{{\tilde L}}\sin{\alpha_1},
\end{equation}
or, in Cartesian coordinates:
\begin{equation}
C^{(3)}_{0ij}=\frac{1}{{\tilde L}}\epsilon_{ijk}X^k.
\end{equation}
The gravitons carry momentum along the $\phi$ direction, as in the previous
section, so their effective action is described by the same action
(\ref{MwavesBI}) with $\phi$ playing the role of the isometric direction. In this
case however there is a contribution from the Chern-Simons action, given that the
3-form potential is electric. Indeed, the Chern-Simons action contributes with a
coupling:
\begin{equation}
S^{{\rm CS}}=i T_0 \int d\tau \ {\rm STr}~\Bigl\{ P[(\incl_X\incl_X)C^{(3)}]\Bigr\}=
\int d\tau \ \frac{2NT_0}{\sqrt{N^2-1}}\frac{r^3}{{\tilde L}}.
\end{equation}
The Born-Infeld part of the action is in this case simpler to compute, given that
$C^{(3)}_{\phi ij}=0$. We have found:
\begin{equation}
S^{{\rm BI}}= -\frac{NT_0}{L}\int d\tau \ \STr \Bigl\{ 
\sqrt{(1+\frac{r^2}{{\tilde L}^2})(\unity + \frac{4L^2r^2}{N^2-1}X^2)} \Bigr \}. 
\end{equation}
Using the same approximation (\ref{STr-Tr}) of the previous section we get:
\be
{\rm STr}\{(X^2)^n\} \approx  r^{2n} N + {\cal O} (\tfrac{1}{N^2-1}),
\ee
and the potential can be written up to terms of order $(N^2-1)^{-3}$ as
\begin{equation}
\label{potmicr2}
V=\frac{NT_0}{L}\sqrt{(1+\frac{r^2}{{\tilde L}^2})(1+\frac{4L^2r^4}{N^2-1})}
-\frac{2NT_0}{\sqrt{N^2-1}}\frac{r^3}{{\tilde L}}.
\end{equation}
This potential have minima at $r=0$ and at
\begin{equation}
\label{solmicr2}
r=\frac{\sqrt{N^2-1}}{2L{\tilde L}},
\end{equation}
which correspond to the point-like graviton and a giant graviton respectively. Both
solutions have the same energy, $E=P_\phi/L$, associated to a massless particle 
with angular momentum $P_\phi=NT_0$. Note however that in this case there is no 
upper bound for the angular momentum, given that the gravitons expand in the $AdS$ 
part of the geometry.

\subsection{The Abelian calculation}

The description of dual giant gravitons in $AdS_4\times S^7$ in terms of the
expanding M2-brane was given in \cite{GMT}. We summarise now these results.

In the Abelian description we have a spherical M2-brane with world volume on the
classical sphere with radius $r$ contained in the $AdS$ part of the geometry, and
carrying angular momentum along $\phi$, as in the previous case. The quantisation
condition (\ref{quanti}) is now:
\begin{equation}
\label{quanti2}
4\pi T_2=\frac{{\bar N}}{L{\tilde L}^2},
\end{equation}
in terms of another integer ${\bar N}$. The Hamiltonian reads \cite{GMT}:
\begin{equation}
\label{potmacr2}
H=\frac{1}{L}\Bigl[ \sqrt{(1+\frac{r^2}{{\tilde L}^2})(P_\phi^2+
\frac{{\bar N}^2r^4}{{\tilde L}^4})}-\frac{{\bar N}r^3}{{\tilde
L}^3}\Bigr].
\end{equation}
The stable solutions correspond in this case
to $r=0$ and to $r=P_\phi {\tilde L}/{\bar N}$,
both of which have energy $E=P_\phi / L$. Using that $P_\phi=NT_0$ and
$T_0={\bar N}/(2L{\tilde L}^2)$, from the quantisation
condition above, we can write
\begin{equation}
r=\frac{P_\phi {\tilde L}}{{\bar N}}=\frac{N}{2L{\tilde L}},
\end{equation}
which is just the large $N$ limit of the microscopical solution given by
(\ref{solmicr2}). Moreover, we can rewrite the Hamiltonian (\ref{potmacr2}) as:
\begin{equation}
H=\frac{NT_0}{L}\sqrt{(1+\frac{r^2}{{\tilde L}^2})(1+\frac{4L^2r^4}{N^2})}
-\frac{2T_0r^3}{{\tilde L}},
\end{equation}
which again agrees with the large $N$ expansion of the 
microscopical potential given by (\ref{potmicr2})!

\sect{Discussion}

We have derived a non-Abelian action for multiple coinciding gravitational waves
in eleven dimensions from a matrix description of massless particles in ten 
dimensions.  We have shown that the non-Abelian couplings to the form fields in this 
action give rise to stable dielectric configurations of gravitons expanding into 
spherical M2-branes. 

By demanding that the non-Abelian Born-Infeld action for multiple 
D0-branes is recovered after compactification, we have been able to extend the validity 
of the eleven-dimensional Born-Infeld action beyond the linearised approximation
of the construction, and give a closed expression valid for more general backgrounds.  
This enabled us to construct configurations of multiple gravitons expanding into
fuzzy dielectric M2-branes, living in the spherical part of $AdS_7\times S^4$ and
the $AdS$ part of $AdS_4\times S^7$. 
We showed that these solutions are in fact the microscopic description of the 
Abelian giant gravitons and dual giant gravitons of \cite{GST, GMT}: physical 
quantities computed in the non-Abelian description, such as the energy of the 
configuration, the radius of the spherical M2-brane and the upper bound for the 
angular momentum, coincide in the large $N$ limit, with $N$ the number of gravitons,
 with the values for these 
quantities computed in the Abelian description. The picture is that in the large $N$
limit the fuzzy two-sphere becomes more and more a better approximation to the 
classical two-sphere from the Abelian description. This idea is not new. It is well 
known that the dielectric effect for D-branes and fundamental strings can be studied
at different, complementary levels, which coincide for large numbers of coinciding 
branes \cite{Emparan, Myers}. It had been suggested before \cite{DTV, Yolanda, BMN, 
JL} that the giant graviton configurations studied in the literature should be the 
large $N$ limit description of some kind of dielectric effect for gravitational waves.
With the 
dielectric solutions to our non-Abelian action we have shown that this is indeed 
the case, hereby completing and connecting the several descriptions of dielectric 
gravitational waves and giant gravitons. 

Moreover, using our action we can find the solution for finite $N$. Take for instance
the potential given by (\ref{potexact}), for the $AdS_7\times S^4$ background.
Expanding the square root to first order, the approximation taken in \cite{Myers},
which in our case is valid when $L^2\sin^2{\theta}<<\alpha^\prime \sqrt{N^2-1}$,
we find:

$$V(\theta)\approx \frac{NT_0}{L\cos{\theta}}(1-
\frac{2L^3\sin^3{\theta}}{\sqrt{N^2-1}}+\frac{2L^6\sin^4{\theta}\cos^2{\theta}}
{N^2-1})\, ,$$

\noindent which has two minima, at $\sin{\theta}=0$ and $\sin{\theta}=\sqrt{N^2-1}
/(2L^3)$, both with an energy $E=NT_0/L$. If we take now the large $N$ limit we
find perfect agreement with the Abelian results.

A few comments are in order. First, a natural question to ask is what is the 
interpretation of the eleven-dimensional linear matrix action (\ref{lin11}) and what
is its relation to other eleven-dimensional linear actions, known from the context
of Matrix theory. In \cite{KT, TvR} a Matrix action with linear couplings to the 
eleven-dimensional background fields was given, which upon reduction over a 
light-like direction yields in ten-dimensions Matrix theory with linear background 
fields. It was shown in \cite{TvR2} that this eleven-dimensional Matrix theory 
action is related via the Seiberg-Sen limit \cite{Sen, Seiberg} to another 
eleven-dimensional matrix action, that gives the linear action for multiple 
D0-branes upon compactification over a space-like circle. The linear action 
(\ref{lin11}) of this paper can be identified with the latter matrix 
action, given that it indeed reduces to the action of multiple D0-branes when 
compactified over the propagation direction of the gravitational waves, as shown 
in the paper. The Seiberg-Sen boost is in fact the coordinate transformation 
necessary to go from the light-cone frame of Matrix theory to the static gauge of 
the above action. 

A second comment we would like to make is that with the closed expression for the 
Born-Infeld action given by (\ref{MwavesBI}), it should be possible to provide a 
microscopical description for other giant graviton configurations, like those in
which the gravitons expand in an $S^5$ living in the spherical part of 
$AdS_4\times S^7$ or in the $AdS$ part of $AdS_7\times S^4$. We should be able to
describe as well, through a chain of dualities, Type IIB gravitons in $AdS_5\times S^5$,
where now the gravitons expand into a D3-brane wound 
around a (fuzzy) three-sphere in the $S^5$. Therefore a different Ansatz from the 
ones we used here will be needed, possibly related to the generators of odd 
fuzzy spheres given in \cite{GR, Ramgoolam}. We leave this for future
investigation.

\vspace{1cm}
\noindent
{\bf \large Acknowledgements}\\
We wish to thank Dominic Brecher, Dave Page and Sanjaye Ramgoolam 
for the useful discussions, and especially Douglas Smith for pointing out some errors
in the previous version. 


\end{document}